\documentclass[%
 amsmath,amssymb,
preprint,%
]{revtex4-1}

\usepackage{graphicx}
\usepackage{dcolumn}
\usepackage{bm}
\usepackage{tikz}
\usepackage{adjustbox}
\usetikzlibrary{arrows, hobby}
\usepackage{wrapfig}
\usepackage{graphicx}
\usepackage{amsmath}
\usepackage{amssymb}
\usepackage{hyperref}
\usepackage{ mathrsfs }
\usepackage{indentfirst}
\usepackage{braket}
\usepackage{color}
\usepackage{empheq}
\usepackage{lipsum,adjustbox}
\usepackage[outdir=./]{epstopdf}
\usetikzlibrary{calc,positioning}

\tikzstyle{beamsplitter}=[fill=blue, fill opacity=0.2]

\usepackage[utf8]{inputenc}
\usepackage[T1]{fontenc}
\usepackage{mathptmx}
\usepackage{etoolbox}

\makeatletter
\def\@email#1#2{%
 \endgroup
 \patchcmd{\titleblock@produce}
  {\frontmatter@RRAPformat}
  {\frontmatter@RRAPformat{\produce@RRAP{*#1\href{mailto:#2}{#2}}}\frontmatter@RRAPformat}
  {}{}
}%
\makeatother
\begin{document}

\preprint{AIP/123-QED}

\title[Quantum effects in rotating reference frames]{Quantum effects in rotating reference frames}
\author{S. P. Kish}
 \affiliation{Department of Quantum Science, Research School of Physics \& Engineering, \\ The Australian National University, Canberra, ACT, Australia.}
 \affiliation{ Centre for Quantum Computation and Communication Technology, \\
	School of Mathematics and Physics, University
	of Queensland, Brisbane, Queensland 4072, Australia
}%
\author{T. C. Ralph}%
\affiliation{ Centre for Quantum Computation and Communication Technology, \\
	School of Mathematics and Physics, University
	of Queensland, Brisbane, Queensland 4072, Australia
}%

\date{\today}

\begin{abstract}
We consider the time delay of interfering {\it single photons} oppositely travelling in the Kerr metric of a rotating massive object. Classically, the time delay shows up as a phase difference between coherent sources of light. In quantum mechanics, the loss in visibility due to the indistinguishability of interfering photons is directly related to the time delay. We can thus observe the Kerr frame dragging effect using the Hong-Ou-Mandel (HOM) dip, a purely quantum mechanical effect. By Einstein's equivalence principle, we can analogously consider a rotating turntable to simulate the Kerr metric. We look at the feasibility of such an experiment using optical fibre, and note a cancellation in the second order dispersion but a direction dependent difference in group velocity. However, for the chosen experimental parameters, we can effectively assume light propagating through a vacuum. 
\end{abstract}

\maketitle

\section{Introduction}
Most experiments performed to date can be explained by a classical theory of curved space-time or quantum mechanics in flat space-time. These remain to a large degree mutually exclusive fields of physics. One of the most fundamental questions of physics today concerns the reconciliation between quantum mechanics and Einstein's theory of general relativity. Quantum mechanics in curved space-time has not been as accessible by experiment. Nonetheless, there have been many proposals to test quantum phenomena such as superposition and entanglement in curved space-time \cite{kempf, european}.  For example, Ref. \cite{mzych0, mzych} considers single photons in superposition at different heights in a gravitational potential. These photons will experience a classical phase shift due to time dilation that is proportional to the gravitational acceleration. The same classical phase can be detected by classical light which does not exhibit the quantum mechanical effect of superposition. However, if the photons have a pulse (coherence) time comparable to the time dilation, significant loss of quantum interference occurs between the photon wavepackets. A loss in the visibility is predicted at the output. 

Similarly, we can consider single photons in a large Sagnac interferometer around a rotating massive body described by the Kerr metric. Due to frame dragging, photons co-propagating with the rotation will have a different arrival time compared with photons that counter-propagate with the rotation. Thus, if this time difference is comparable to the pulse time of the photon, loss of quantum interference between photon wavepackets occurs. The loss in visibility now depends on the Kerr parameter $a$. We can consider an analogous system on a rotating turntable. By Einstein's equivalence principle, we might expect to be able to find situations where the physics is equivalent whether they are in the space-time of a Kerr metric or accelerating due to the rotation of the turntable. We can observe the time difference due to light velocity using the Hong-Ou-Mandel (HOM) effect, a purely quantum mechanical effect \cite{enk}. The time delay due to rotation at the single photon level as a phase shift has been observed previously \cite{sag} and also more recently, with a $N=2$ NOON state \cite{fink}. The effect of uniform rotation on the distinguishability of photons was recently observed in a Sagnac-type interferometer on a rotating platform \cite{rest}. An extension of the experiment in Ref \cite{rest} as suggested by the authors is to use satellites in Earth's orbit to perform tests of quantum mechanics at the interface of general relativity (i.e. frame-dragging). Moreover, in a recent paper, frame dragging and the HOM dip has been studied in common-path setup restricted to the surface of the Earth \cite{brady}. However, loss in the visibility of the  quantum interference in the Kerr metric has not been investigated in detail in an analogous rotating frame set-up. 

In this paper we consider the time delay in the Kerr metric for the situation where a stationary observer sends a superposition of a co- and counter-propagating photon half-way around the massive body. We find that there is a visibility loss for Gaussian wavepackets due to the time disjoint space-time of the Kerr metric which agrees with recent work \cite{toros}. Next, we consider a turntable experiment to simulate this effect. We use a single-photon Sagnac interferometer to see the loss in visibility due to the time delay caused by rotation. We also consider the relativistic effects of the optical fibre medium. We find that the second order dispersion cancels out in agreement with previous work \cite{steinberg}, but small effects due to a difference in group velocity remain. Nonetheless, experimental parameters accessible with current technology are proposed for future experiments to observe the visibility loss due to rotation analogous to what would be observed in a single-photon Sagnac interferometer in the Kerr metric of a massive object. Ultimately, one may observe such effects at the interface between quantum mechanics (via single-photon interference) and non-trivial rotating space-times. 

\section{Kerr metric}
Although the Hartle-Thorne metric describes the exterior metric of a massive object with a mass quadrupole moment $q$, the approximate Kerr metric in the weak field limit disregards $q$ which is of second order in angular momentum. Therefore, planets or stars can be described by the Kerr metric up to first order in the angular momentum. For a rotating black hole $q=0$ and the Kerr metric describes the effect of the dragging of the space-time. The Kerr line element in Boyer-Lindquist coordinates $(t,r,\theta,\phi)$ is \cite{visser}

\begin{equation}
\begin{split}
ds^2&=-(1-\frac{r_s r}{\Sigma})dt^2+\frac{\Sigma}{\Delta} dr^2+\Sigma d\theta^2 \\
&+(r^2+a^2+\frac{r_s r a^2}{\Sigma} \sin^2{\theta})\sin^2{\theta} d\phi^2-\frac{2 r_s r a \sin^2{\theta}}{\Sigma} d\phi dt,
\end{split}
\end{equation}

where $\Delta:=r^2-r_s r+a^2$, $\Sigma:=r^2+a^2 \cos^2{\theta}$ and $a=\frac{J}{M c}$ where $J$ is the angular momentum of the object of mass $M$. Note that the Schwarzschild radius is $r_s=\frac{2GM}{c^2}\equiv 2 M$ where we work in natural units for which $c=1$ and $G=1$. 
In the equatorial plane of the metric ($\theta=\frac{\pi}{2}$), a tangential velocity of light can be obtained in terms of tangential proper distance per Kerr coordinate time, solved by setting $ds^2=0$. The full solution is given by (see also Ref. \cite{kish})

   \begin{equation}
   \begin{split}
    \frac{d x}{dt}&=\frac{r_s a}{r \sqrt{r^2+a^2(1+r_s/r)}} \\
    &\pm \sqrt{\frac{ r_s^2 a^2}{r^2(r^2+a^2(1+r_s/r))}+(1-\frac{r_s}{r})}.
    \end{split}
    \label{full}
   \end{equation}

However, if $\frac{a}{r}<<1$ and $\frac{r_s}{r}<<1$ we have the weak field solution valid for massive planets
   
     \begin{equation}
     \begin{split}
     \frac{d x}{dt} &\approx \frac{r_s a}{r^2} \pm \sqrt{1-\frac{r_s}{r}} \\
     &\approx \pm (1-\frac{r_s}{2 r} \pm \frac{r_s a}{r^2}),
     \end{split}
     \label{dxdt}
     \end{equation} where we have used the Taylor expansion $\sqrt{1-x}\approx 1-\frac{x}{2}$. We have two solutions representing counter- and co-rotating light. 
\subsection{Kerr phase difference}
According to Eq. (\ref{dxdt}), the speed of light as seen by a far-away observer in the weak field limit is $c_A=(1-\frac{r_s}{2 r} + \frac{r_s a}{r^2})$ and $c_B=-(1-\frac{r_s}{2 r} - \frac{r_s a}{r^2})$ for co- and counter- propagating light, respectively. Hence, the phase difference between co- and counter- propagating single photon wavepackets is

\begin{equation}
\begin{split}
\Delta \Phi&=\Phi_B -\Phi_A= \omega L ( \frac{1}{c_B}-\frac{1}{c_A}) \\
&\approx \omega L (\frac{1}{1-\frac{r_s}{2 r} + \frac{r_s a}{r^2}}\\
&-\frac{1}{1-\frac{r_s}{2 r} - \frac{r_s a}{r^2}}) \\
&\approx 2 \omega L \frac{r_s a}{r^2} (1+\frac{r_s}{r}) \approx 2 \omega \pi \frac{r_s a}{r}.
\end{split}
\label{kerrphi}
\end{equation}

Note that we've neglected second order terms in $\frac{a}{r}$ and above, since $r>>a$ in the weak field limit. For Earth's parameters $a=3.9$ $m$, $r_s=9$ $mm$ on the surface $r=6.37 \times 10^7$ $m$ with visible light of frequency $\omega=k=2 \times 10^6$ $m^{-1}$, the magnitude of the classical Kerr phase half way around the Earth is $\Delta \Phi_{Kerr}=\omega \Delta t_{Kerr} \approx 7 \times 10^{-3}$ where $\Delta t_{Kerr}=2 L \frac{r_s a}{r^2}$. The time-delay due to the effect of the Kerr metric derived here agrees with Ref. \cite{rest}. In principle, we could use a large classical Sagnac interferometer to measure this phase.  
\section{Single photon Sagnac interferometer in the Kerr metric}

We consider a thought experiment where a single photon is in a superposition of two paths around the rotating planet. A stationary single photon source hovering above the rotating planet releases a photon that passes through a beamsplitter and forms a superposition of the paths $A$ and $B$ with phases $\Phi_A=\omega t_A$ and $\Phi_B=\omega t_B$. Paths $A$ and $B$ move semi-circularly around the planet co- and counter- propagating, respectively, with the rotation direction of the planet. These recombine at a second beamsplitter that is half-way around the planet and are detected by a photon number counter. 

We begin with the initial state $\ket{1} \ket{0}$ passing through a beamsplitter

\begin{equation}
\ket{1} \ket{0} \rightarrow \frac{1}{\sqrt{2}}(\ket{1} \ket{0}+i\ket{0} \ket{1}).
\end{equation}

The two arms experience a different phase shift depending on their trajectories. If the photon is moving with the rotation, it will acquire the phase $\Phi_A$ and against the rotation, it will acquire the phase $\Phi_B$. Thus we have the state after the first beamsplitter and subsequent propagation to the second beamsplitter

\begin{equation}
 \frac{1}{\sqrt{2}}(e^{i\Phi_A}\ket{1} \ket{0}+ie^{i\Phi_B}\ket{0} \ket{1}).
\end{equation}

After the second beamsplitter, the state becomes

\begin{equation}
 \frac{1}{2}(e^{i\Phi_A}(\ket{1} \ket{0}+i\ket{0}\ket{1}+ie^{i\Phi_B}(i\ket{0} \ket{1}+\ket{1}\ket{0})).
\end{equation}

Thus the number of photons at the output of the second beamsplitter is

\begin{equation}
\begin{split}
 \braket{N}&=\bra{0}\bra{0} \frac{1}{2}(e^{-i\Phi_A}-ie^{-i\Phi_B})\frac{1}{2}(e^{i\Phi_A}+ie^{i\Phi_B})\ket{0}\ket{0}\\
 &=\frac{1}{2} (1+\sin{(\Phi_A-\Phi_B)}).
 \end{split}
\end{equation}

This particular phase can also be measured using classical coherent probe states. Due to this reason, it's usually referred to as a ``classical phase". 

Consider the single photon mode distribution $f(\omega) a^\dagger_\omega \ket{0}$ where 
\begin{equation}
f(\omega)=(\frac{1}{\pi \sigma^2})^{1/4} \exp{(-\frac{1}{2 \sigma^2}(\omega-\omega_0)^2)},
\end{equation}
is the Gaussian distribution with centre frequency $\omega_0$ and $\sigma$ is the pulse width in frequency space.  Thus at the output we have
\begin{equation}
\begin{split}
 \braket{N}&=  \int{d \omega} |f(\omega)|^2 \braket{a^\dagger_\omega a_\omega} \\
 &=\frac{1}{2}(1+\int{d\omega (\frac{1}{\pi\sigma^2})^{1/2} \exp{(-\frac{1}{\sigma^2} (\omega-\omega_0)^2)\sin{\omega \Delta t}}}) \\
 &=\frac{1}{2}(1+e^{-(\frac{\Delta \Phi \sigma}{\omega_0})^2}\sin{\Delta \Phi}),
 \end{split}
\end{equation}
where the integral is over all positive frequencies $\omega$ and $\Delta \Phi$ is given by Eq. \eqref{kerrphi} in the Kerr metric. The visibility is given by 
\begin{equation}
\mathcal{V}=e^{-(\frac{\Delta \phi \sigma}{\omega})^2}=e^{-(\Delta t \sigma)^2}
\label{vis}
\end{equation}

The additional visibility loss is classified as a quantum effect because it is due to the interference of wavefunctions as opposed to interference of classical modes of the field had we used coherent probe states.

\subsection{Two-way velocity of light}
We now want to determine the average velocity of light that returns to the stationary observer. We expect this to be isotropic and equal to $c=1$. We now consider a double-sided mirror at halfway around the rotating planet that reflects both of the photons back to the original source and the photons interfere. The phase of the co-propagating photon $A$ before interfering with the counter-propagating photon $B$ is 
\begin{equation}
\begin{split}
\Phi_A+\Phi_A'&=\omega L (\frac{1}{1-\frac{r_s}{2 r} + \frac{r_s a}{r^2}})+\omega L (\frac{1}{1-\frac{r_s}{2 r} - \frac{r_s a}{r^2}})\\
&\approx \omega L (1+\frac{r_s}{2r}),
\end{split}
\end{equation} where $\Phi_A'=\Phi_B$ since this the counter-propagating phase. Note we have used the weak field approximation. The mean velocity of the light as seen by a far-away observer is thus $c_{mean}=\frac{1}{1+\frac{r_s}{r}}\approx 1-\frac{r_s}{r}$. Locally, the velocity of light is $\frac{dx}{d\tau}=c_{mean} \frac{dt}{d\tau}=1$ where $\frac{dt}{d\tau}=(1-\frac{r_s}{r})^{-1}$ from setting $dr=0, d\phi=0, d\theta=0$ in the Kerr metric. This implies that the local velocity of light is $c=1$ and isotropic as expected.

Similarly, for the initially counter-propagating photon we have the same phase. Thus no phase difference is detected and the observer infers that $c=1$. 

\subsection{Visibility loss in the Kerr metric of Earth}
Assuming that the mass quadrupole moment $q$ is negligible, the approximate metric in the weak field limit is the Kerr metric. Therefore, if we consider the massive planet to be Earth, the two photon paths will undergo a time delay due to the frame dragging caused by the Kerr metric.  
Therefore, we have $\mathcal{V}=e^{-(\Delta t_{Kerr} \sigma)^2}e^{-(\frac{2 \pi r_s a \sigma}{r})^2}\approx 1-1.5 \times 10^{-10}$ where $a=3.9$ $m$, $r_s=0.009$ $m$ and $r=6.37\times 10^6$ $m$. The visibility loss in the quantum interference would be far too small to be detected on Earth's surface. The radial position at which the visibility would be significant is $r=2 \pi r_s a \sigma\approx 800$ $m$, assuming a neutron star with Earth's mass.

\subsection{Extremal black holes}
\begin{figure}
\centering
\includegraphics[scale=0.6]{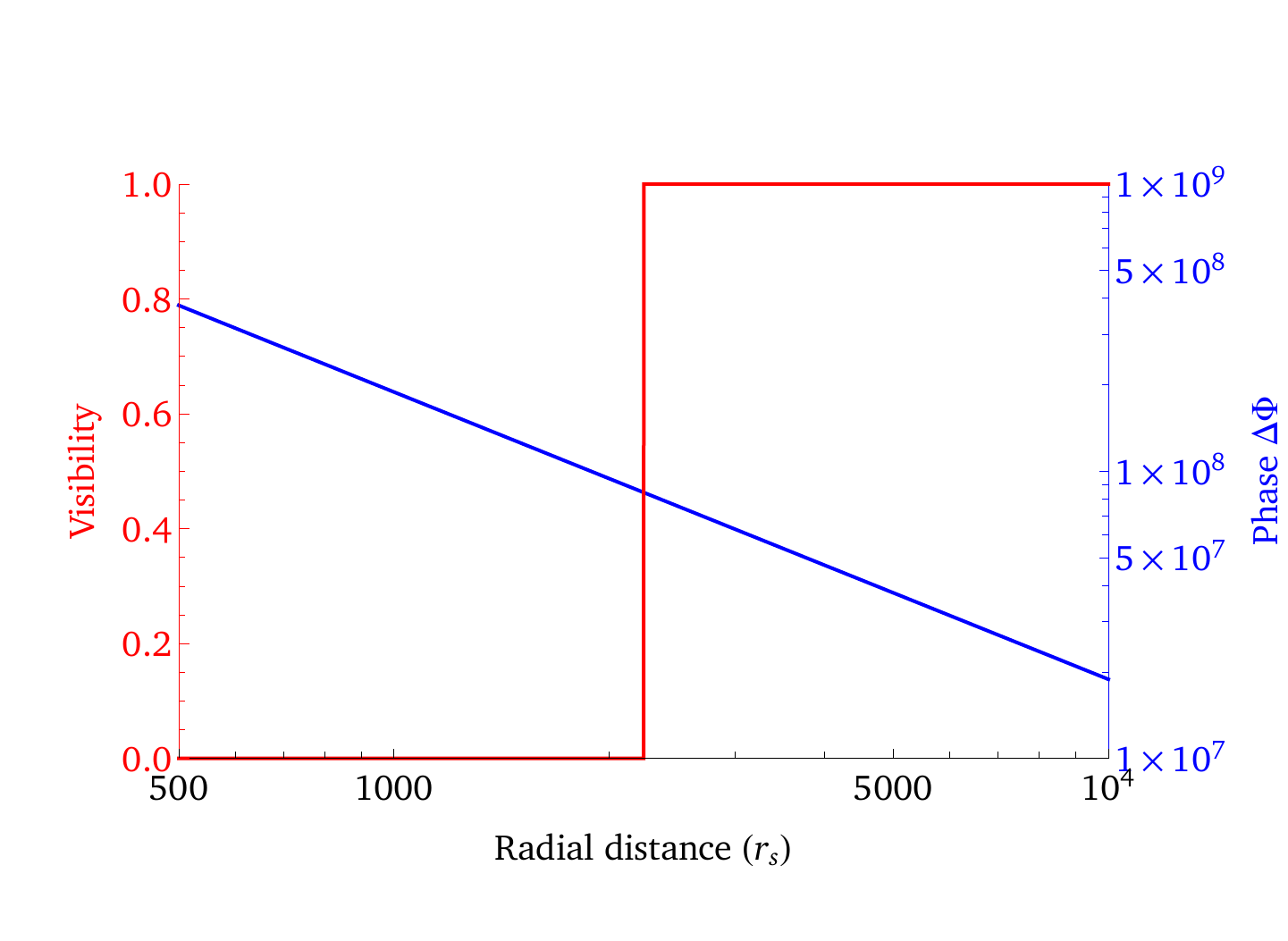}
\caption{Phase difference (blue) and visibility (red) of single photons plotted against radial position in units of the Schwarzschild radius $r_s$ from centre of a rotating black hole. (Note the extremal black hole parameters $r_s=30$ $km$, $a'=\frac{1}{4}$ and $\sigma=3.5 \times 10^3$ $m^{-1}$.)}
\label{graph}
\end{figure}

An extremal black hole has a minimum possible mass for a given angular momentum. We consider the full solution to the speed of light without any weak field approximations for a near extremal black hole where the black hole mass is a few orders above the angular momentum. For a black hole, due to symmetry, the quadrupole mass moment is $q=0$ and the space-time is described by the Kerr metric. The phase is therefore

\begin{equation}
\Delta \Phi=\omega (t_B-t_A)=k L (\frac{1}{c_B}-\frac{1}{c_A}),
\label{bh}
\end{equation}

where $c_{A,B}=\frac{r_s a}{r_B \sqrt{r_B^2+a^2(1+r_s/r_B)}} \pm \sqrt{\frac{r_s^2 a^2}{r_B^2(r_B^2+a^2(1+r_s/r_B))}+(1-\frac{r_s}{r_B})}$. Using units of $r_s$, $a\rightarrow a' r_s$, $r_A \rightarrow r'_A r_s$ and $r_B \rightarrow r'_B r_s$. This simplifies to $c_{A,B}=\frac{1}{r'_B \sqrt{ \frac{r_B'^2}{a'^2}+(1+1/r'_B)}} \pm \sqrt{\frac{1}{r_B'^2((\frac{r_B'^2}{a'^2}+(1+1/r'_B))}+(1-\frac{1}{r'_B})}$. We consider the values of a near extremal black hole with angular momentum parameter $a$ on the order of $r_s$ with $r_s=30$ $km$, $a'=\frac{1}{4}$, interferometer length of $L=2 \pi r$ and $\sigma=3.5 \times 10^3$ $m^{-1}$. In Fig. \ref{graph}, we have the phase and the visibility plotted against the radial distance away from the black hole centre. At a particular radial distance, the visibility loss is significant and the wavepackets become completely separated very near the event horizon. 

\section{Rotating reference frame}
\begin{figure}

    
\includegraphics[scale=0.99]{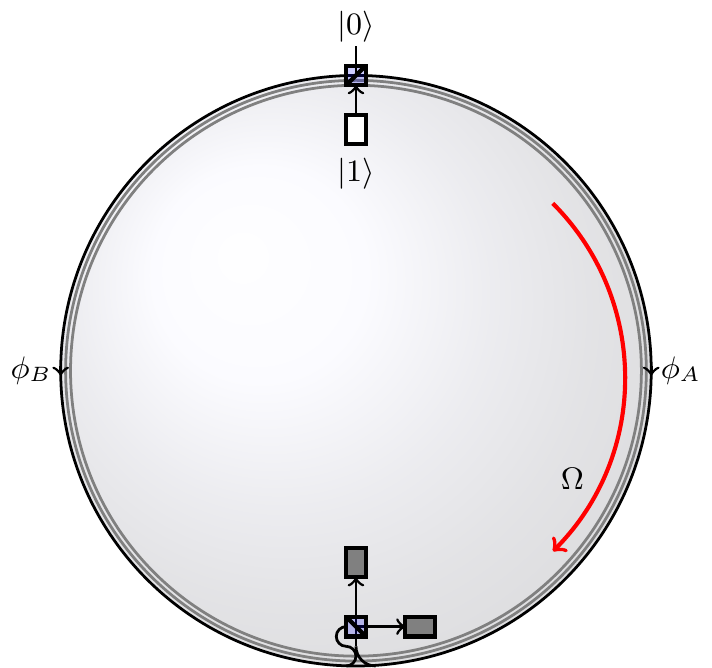}

\centering
\caption{A HOM Sagnac interferometer on a turntable rotating at angular frequency $\Omega$. A photon source (white) releases two photons that pass through a beamsplitter and forms a superposition of the paths $A$ and $B$ with phases $\Phi_{A}$ and $\Phi_B$. These recombine at the second beamsplitter and are detected by a photon counter (gray). }
\label{turntable}
\end{figure}

In this section, we will compare the metric of an inertial observer in a rotating reference frame with an observer in the Kerr metric. Consider an inertial observer on a rotating turntable at a constant radius $r$ with angular frequency $\Omega$ as depicted in Fig. \ref{turntable}. 

The metric for a rotating observer in the (1+1) space-time is given by transforming the flat metric using $d \phi \rightarrow d\phi + \Omega dt$ 

\begin{equation}
ds^2_{Rotation}=dt^2-r^2 d\phi^2 \rightarrow (1-v^2) dt^2-2 v r_t dt d\phi-r_t^2 d\phi^2,
\label{rotating}
\end{equation}
where the $t$ and $\phi$ coordinates are the inertial coordinates in the rotating metric and where $v=\Omega r_t$ is the tangential velocity. We compare the rotating metric with the Kerr metric at constant radius in a (1+1) dimensional space-time
\begin{equation}
ds^2_{Kerr}=(1-\frac{r_s}{r}) dt^2-\frac{2 r_s a}{r} dt d\phi-r^2 d\phi^2.
\end{equation}
We require a coordinate transformation for it to match with equation \ref{rotating}. Evidently, we require that $r d\phi_{Kerr}=r_t d\phi_{Rotation}$ implying the measured tangential distances are the same. We set the size of the turntable $r_t=r$ which implies that the angular coordinates are equivalent $d\phi_{Kerr}=d\phi_{Rotation}$. However, we make the following time coordinate transformation in the Kerr metric
\begin{equation}
dt\rightarrow \sqrt{\frac{1-v^2}{1-\frac{r_s}{r}}} dt.
\label{timetransform}
\end{equation}
Thus
\begin{equation}
\begin{split}
ds^2_{Kerr}=(1-v^2)dt^2-\frac{2 r_s a}{r} \sqrt{\frac{1-v^2}{1-\frac{r_s}{r}}} dt d\phi-r^2 d\phi^2.
\end{split}
\end{equation}
The $g_{t \phi}$ components must match and we have to solve for $v$ in the equation $v r=\frac{r_s a}{r} \sqrt{\frac{1-v^2}{1-\frac{r_s}{r}}}$. We find that the tangential velocity of the rotating turntable for simulating a Kerr reference frame must be

\begin{equation}
v=\pm \frac{r_s a}{r^2 \sqrt{1-\frac{r_s}{r}+\frac{r_s^2 a^2}{r^4}}}\approx \pm \frac{r_s a}{r^2 \sqrt{1-\frac{r_s}{r}}}.
\label{tv}
\end{equation}
By Einstein's equivalence principle, the local observer cannot tell whether they are in the (1+1) local space-time of a Kerr metric or in an inertial rotating reference frame of a turntable. Therefore, we can simulate tangential motion in a (1+1) Kerr space-time if we choose an appropriate tangnetial velocity for the turntable, and use appropriate clocks. 

For example, for a turntable of Earth's radius $r=6.37 \times 10^7$ $m$, we can simulate the Earth's Kerr metric with a turntable tangential velocity of $v=\frac{r_s a}{r^2 \sqrt{1-\frac{r_s}{r}}} \approx 2.6 \times 10^{-7}$ $m/s$ or angular frequency $\Omega=4.1\times 10^{-14}$ $rad/s$. Alternatively, for an Earth mass black hole $r_s=0.009\text{ m}$, and turntable radius of $r=100$ $m$ then the tangential velocity of the turntable with the same radius would have to be $v=110$ $m/s$ to match the effects of the Kerr metric. On the other hand, for a $10$ solar mass black hole ($r_s=30\text{ km}$) rotating at $a=r_s/100$ with an observer $10 r_s$ distance from the centre, the tangential velocity of the turntable at the same radius is $v=30,000 \text{ m/s}$. We note that these high tangential velocities are not currently experimentally feasible.

%
%

\subsection{Time shift}
An alternative scenario is where both observers in either metrics use their respective coordinate times but we require the time shifts imposed on tangentially propagating light rays to be equal. We consider the perspective of an inertial observer in the laboratory frame observing the rotating platform that sends light for a round trip. The round trip for the light as seen by the inertial observer is obtained by considering the distance travelled by the co-moving light $L+v t= t$ where $t$ is the total time of the round trip. Thus

\begin{equation}
\Delta t=\frac{L}{1-v}-L=\frac{2 \pi r_t}{\sqrt{1-v^2}(1-v)}-\frac{2 \pi r_t}{\sqrt{1-v^2}},
\end{equation}
where $r_t$ is the turntable radius as seen by the inertial observer in the laboratory frame (note we have accounted for the time dilation). In the Kerr metric, as seen by a far-away observer the time of the round trip (w.r.t. to the stationary observer) around a massive planet is (we have used Eq. \eqref{dxdt})

\begin{equation}
\begin{split}
\Delta t_{Kerr}&\approx\frac{2 \pi r }{1-\frac{r_s a}{r^2}-\frac{r_s}{2 r}}-\frac{2 \pi r}{1-\frac{r_s}{r}}\\
&\approx 2 \pi r(1+\frac{r_s}{2 r}+\frac{r_s a}{r^2})-2 \pi r (1+\frac{r_s}{2r})= \frac{2 \pi r_s a}{r},
\end{split}
\end{equation}
where $r$ is the radius in the Kerr metric.
Equating these two far-away times ($\Delta t=\Delta t_{Kerr}$) and without assuming $v$ is small, we have
%

\begin{equation}
\frac{r_s a}{r r_t}=\frac{1}{\sqrt{1-v^2}(1-v)}-\frac{1}{\sqrt{1-v^2}}=\frac{v}{\sqrt{1-v^2}}.
\end{equation}

Thus we have the velocity
\begin{equation}
v= \frac{r_s a}{r r_t \sqrt{1+\frac{r_s^2 a^2}{r^4}}} \approx \pm \frac{r_s a}{r r_t}
\end{equation}
For example, for a turntable of radius $r_t=0.2$ $m$ and $r=r_E$ equal to the Earth's radius we have a turntable velocity of $v=\pm 0.8$ $m/s$ or angular frequency $\Omega=4$ $rad/s$. Compared to the previous example of an Earth size turntable, the accumulated effect on the smaller turntable is smaller and thus the velocity must be larger to compensate. Various black-hole scenarios could be simulated using lab-sized turn-tables using this method. 
We note that if the turntable radius is equal to the Kerr metric radius $r_t=r$ and we use the metric times in Equations \ref{rotating} and \ref{timetransform}. We obtain the equation

\begin{equation}
\frac{r_s a}{r^2} \sqrt{\frac{1-v^2}{1-\frac{r_s}{r}}}=v,
\end{equation} which, solving for $v$ is equivalent to Eq. \eqref{tv}. Therefore, we have proven that a transformation of time coordinates isn't necessary to quantitatively simulate the Kerr metric with a rotating turntable. 

%
%
%

\subsection{Phase}
For a rotating reference frame, we solve for the light null geodesic to obtain the tangential velocity of light $\dot{x}=r\frac{d \phi}{dt}$. Setting $ds^2=0$ we have $(1-v^2) - 2 v \dot{x}-\dot{x}^2=0$. We thus have two solutions $\dot{x}=1\pm v$. The cross term component $dt d\phi$ once again is the cause of the anisotropy of light. With the rotation of the turntable, the phase of the photon is $\Phi_A=\omega \frac{L}{c_A}=\omega \frac{L}{1+v}$ and against the rotation, with the phase $\Phi_B=\omega \frac{L}{c_B}=\omega \frac{L}{1-v}$. Thus the phase difference for the turntable is
\begin{equation}
\Delta \Phi_{Rotation}=\frac{2 \omega v L}{1-v^2},
\label{rotphase}
\end{equation}

which is the same phase obtained by a classical Sagnac interferometer.

\subsection{Minimum velocity for significant visibility loss}
We substitute the phase in Eq. \eqref{rotphase} in Eq. \eqref{vis} to obtain the visibility
\begin{equation}
\mathcal{V}=e^{-\sigma^2  \frac{\Delta \phi^2}{\omega^2}}=e^{-4 \frac{v^2 L^2 \sigma^2}{(1-v^2)^2}}=e^{-4 \frac{v^2} {(1-v^2)} \pi^2 r^2 \sigma^2}.
\end{equation}

Note that $L$ is the total contracted length travelled by the light. For the case where the photons meet half way we have $L=\pi r \sqrt{1-v^2}$. For significant visibility loss the velocity needed for the rotating platform is given by solving $4 \frac{v^2}{ (1-v^2)} \pi^2 R^2 \sigma^2=1$. Rearranging, 
\begin{equation}
v_{min}=\frac{1}{\sqrt{4 \pi^2 r^2 \sigma^2+1}} \approx \frac{1}{2 \pi r \sigma}.
\label{min}
\end{equation}


We consider $\frac{\sigma}{c}=\frac{1}{\Delta t_p c}=3.3 \times 10^3$ $m^{-1}$ corresponding to picosecond pulses and a rotating platform of radius $r=5$ $m$. Thus $v\approx \frac{1}{2 \pi R \sigma}\approx 2900 m/s$. If smaller pulses of $100$ femtoseconds are used then $v \approx 290 m/s$. The amount of g-force for this velocity is $1700$ $g$. However, we can simply increase $L$ by increasing the number of windings around the turntable, thus accumulating the effect, and making it significant for much lower velocities. Although this is not feasible without increasing the fibre windings, we discuss and provide a reasonable optical fibre length using a coherent light source in Section \ref{coherence}.


%
%

\section{Two photons input (HOM interference)}
In our thought experiment, we considered a superposition of a single photon travelling two paths. The probability of detection depends on the classical phase shift $\Delta \Phi$, and also the visibility of the quantum interference. The Hong-Ou-Mandel (HOM) interferometer uses two photons as input. However, in this setup, a physical time delay between the paths is explicit in the probability of detection. The loss of interference is due to loss of indistinguishability of the photons and has no classical analogue. The HOM effect can be interpreted as more quantum due to this strong quantum interference. 
We consider a source of single photons travelling paths $A$ and $B$ to a beamsplitter. The state after the beamsplitter becomes

\begin{equation}
\begin{split}
a^\dagger_1 a^\dagger_2 e^{i \omega \Delta t} \ket{0} \ket{0} &\rightarrow \frac{1}{2} (a^\dagger_3+ia^\dagger_4) (a^\dagger_4+ia^\dagger_3) e^{i \omega \Delta t} \ket{0} \ket{0} \\
&=\frac{e^{i \omega \Delta t}i}{2} (\ket{2}\ket{0}+\ket{0}\ket{2}),
\end{split}
\end{equation}

where $e^{i \omega \Delta t}$ is a global phase difference between the two modes. We thus have photon pair incidences at the final output of the beamsplitter. In this case the time delay doesn't affect the coincidence probability, since the photons are not distinguishable. 

Now we consider a source of photons with arbitrary frequency distributions $f(\omega_1)$ and $g(\omega_2)$ sending the initial state $\ket{1} \ket{1}=\int{d \omega_1 f(\omega_1) a_1^\dagger (\omega_1)}\int{d \omega_2 g(\omega_2) a_2^\dagger (\omega_2)}e^{-i \omega_2 \Delta t} \ket{0}\ket{0}$ to a beamsplitter. The transformation is

\begin{equation}
\begin{split}
a^\dagger_1 a^\dagger_2 \ket{0} \ket{0} &\rightarrow \frac{1}{2} \int{d\omega_1 f(\omega_1)} (a^\dagger_3(\omega_1)+i  a^\dagger_4(\omega_1)) \\
&
\times \int{d\omega_2 g(\omega_2)} (a^\dagger_4(\omega_2)+i  a^\dagger_3 (\omega_3)) e^{-i \omega_2 \Delta t} \ket{0} \ket{0} \\
&= \frac{1}{2}  \int{d\omega_1 f(\omega_1)} \int{d\omega_2 g(\omega_2) e^{-i \omega_2 \Delta t}} \\
&\times (i a^\dagger_3 (\omega_1)a^\dagger_3 (\omega_2)+a^\dagger_3 (\omega_1)a^\dagger_4 (\omega_2) \\
&- a^\dagger_4 (\omega_1)a^\dagger_3 (\omega_2)+i a^\dagger_4 (\omega_1)a^\dagger_4 (\omega_2)) \ket{0} \ket{0}.
\end{split}
\end{equation}

The detection probability of a photon pair in either mode is determined by modelling the detectors as having a flat frequency response with the projector $P_3=\int{d\omega a^\dagger(\omega) \ket{0}\ket{0} \bra{0}\bra{0} a(\omega)}$. These calculations have been done in Ref. \cite{bran}, and for photons of the same frequency distribution $f(\omega)=g(\omega)$, the probability is

\begin{equation}
P=\frac{1}{2}-\frac{1}{2} \int{d\omega_1 |f(\omega_1)|^2 e^{-i \omega_1 \Delta t} } \int{d\omega_2 |f(\omega_2)|^2 e^{i \omega_2 \Delta t} }.
\label{hom}
\end{equation}

For photons with Gaussian frequency distribution of pulse width $\sigma$, we evaluate Eq. \eqref{hom} to obtain
\begin{equation}
P_{Gauss}=\frac{1}{2}-\frac{1}{2} e^{-\frac{\sigma^2 \Delta t^2}{2}}.
\end{equation}

For zero time delay $\Delta t=0$, the photons are indistinguishable. At this point, the visibility $\mathcal{V}=\text{Tr}(\rho_a \rho_b)$ of the two photon states $\rho_a$, $\rho_b$ is equal to the purity $\text{Tr}(\rho^2)$ of the two photons $\rho=\rho_b=\rho_a$ \cite{bran}. Thus the visibility, defined as $1-P_0/P_{max}$, where $P_0$ is the coincidence probability at $\Delta t=0$ and $P_{max}$ is the probability for $\Delta t$ larger than the photon coherence length, is $100 \%$ which is known as the Hong-Ou-Mandel dip. The coincidence count drops to zero when the two input photons are completely identical. For the case of the rotating turntable, the photons will become more distinguishable as $\Delta t$ increases. The velocity at which this becomes significant is the same as in Eq. \eqref{min} since the visibility is the same and the time delay is $\Delta t=\frac{4 v L}{1-v^2}$. We can calibrate the dip for $100$ $\%$ visibility using a controlled time delay in one of the arms and vary the rotational velocity of the turntable.

Compared to using a single photon interferometer, the HOM interferometer is based on the indistinguishability of the photons interfering with each other. As in Ref.  \cite{enk}, the HOM effect measures a physical time delay as opposed to a phase shift. 

\subsection{Two-way velocity of light}
Similarly to the Kerr metric, the two-way velocity of light as measured by the inertial observer on the rotating turntable should be isotropic. We now consider a double sided mirror that reflects both of the light beams back to the original source. In this case, our phase difference is $\phi_A'=\phi_A+\frac{\omega L}{c_B}=\omega L (\frac{1}{1+v}+\frac{1}{1-v})= \frac{2 \omega L}{1-v^2}$ but $L=\pi R\sqrt{1-v^2}$. Therefore $\phi'_A=\frac{2 \omega L}{\sqrt{1-v^2}}$. Similarly, for the counter-propagating beam of light $\phi_B'=\phi_B+\frac{\omega L}{c_A}=\omega L (\frac{1}{1-v}+\frac{1}{1+v})= \phi'_A$. Thus the phase difference is zero. In other words, the ``two-way velocity" of light is isotropic and $c=1$. This demonstrates that observers riding on the turntable would also measure $c=1$ between points around the circumference.

\section{Experimental feasibility in optical fibre}
We consider effects specific to the fibre optic set-up and the feasibility of observing the HOM dip on a rotating platform. In a medium, the dispersion relation describes the relation of the frequency $\omega$ to its wavenumber $k$.  The dispersion relation is Taylor expanded as $\omega(k)=\omega(k_0)+(k-k_0) v_g + \frac{1}{2}(k-k_0)^2 \frac{d^2\omega}{dk^2}$ where $v_g=\frac{1}{\alpha}=\frac{d\omega}{dk}$ is the group velocity and $\frac{1}{\beta}=\frac{d^2\omega}{dk^2}$ is the inverse group velocity dispersion. For linear dispersion $\omega=\frac{k}{n}$ implying $\frac{d^2\omega}{dk^2}=0$ and the group velocity $v_g=\frac{1}{n}$ is equal to the phase velocity $v_p$. 

In relativity, the phase velocities of light in a medium in the stationary reference frame of the laboratory transform according to the Lorentz transformations. 

\subsection{Lorentz transformations in a moving medium}
We consider using fibre optic cable to guide the photon half-way around the turntable as seen in Fig. \ref{turntable}.  In a medium, the velocity of light is slowed down by the factor $1/n$ where $n$ is the refractive index. The velocity of light in the medium depends on direction of the rotation relative to the observer in the laboratory reference frame.

According to the velocity composition law, for a moving medium, the velocity of light as seen in the laboratory reference frame of the rotating reference frame is 
\begin{equation}
c_A=\frac{\frac{c}{n}-v}{1-\frac{v}{c n}},
\label{phasev}
\end{equation} 
which is equivalently the phase velocity. Therefore, $L-v t=t \frac{\frac{c}{n}-v}{1-\frac{v}{c n}}$, and $t=\frac{L(n-v)}{1-v^2}$ implying that the effective velocity of light as seen by a far-away observer is $c_A'=\frac{1-v^2}{n-v}$ and similarly, $c_B'=\frac{1-v^2}{n+v}$.

\subsection{Phase velocity}

Using the phase velocity in Eq. \eqref{phasev}, the new phase is therefore $\Delta \Phi =  \omega_0 (\frac{L}{c_A'}-\frac{L}{c_B'}) =\frac{2v}{1-v^2} \omega_0 L $. Coincidentally, the final phase doesn't depend on the refractive index if the fibres are equal length $L$. However, if these are unequal $L_A=L$ and $L_B=L+\Delta L$, then $\Delta \Phi'=\Delta \Phi+\omega_0 \Delta L \frac{n}{1-v^2}$. The HOM dip will measure the time delay $\Delta t'=\frac{4 v L+2 n \Delta L}{1-v^2}$. If $\Delta L$ is on the length scale of the coherence length then we can simply cancel this out with a controlled time delay. 

\begin{figure}
\centering
\includegraphics[scale=0.9]{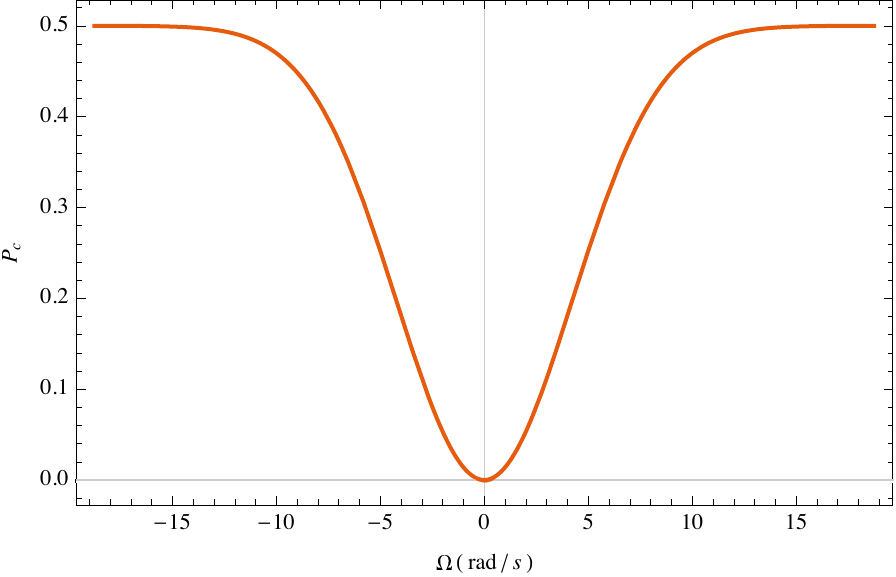}
\caption{Probability of photon detection with angular frequency of turntable for parameters $L=10$ $km$, $R=20$ $cm$ and $\sigma=4000 \pi$.}
\label{graph1}
\end{figure}

As in Fig. \ref{graph1}, a time delay initially sets the visibility to $100 \%$ with the turntable at rest. The turntable is slowly rotated and the visibility decreases to $0 \%$. The HOM dip will be initially centered around $v=0$ where $\Delta t'_0=2 \Delta L n+\Delta t_{Control}=0$. Thus as the turntable is slowly rotated we have
\begin{equation}
\begin{split}
\Delta t'&=\frac{4 v L}{1-v^2}+2\Delta L \frac{n}{1-v^2}-2\Delta L n \\
&\approx \frac{4 v L}{1-v^2} + 2 \Delta L n v^2.
\end{split}
\end{equation}

There is as shift in the centre of the HOM dip. $\Delta L$ depends on the experimental error of the measured fibre lengths and the velocity of the turntable. For example, for a slowly rotating turntable of $\Omega=2 \pi$ $rad/s$ and $R=20$ $cm$ with minimal vibrations, a mismatch in the length of the optical fibres of $\Delta L= 1$ $cm$ would shift the HOM dip by $\Delta t_{Error} = 2 \Delta L n v^2 \approx 3\times 10^{-11}$ $s$ or relative to the leading term $\frac{\Delta t_{Error}}{\Delta t}= \frac{\Delta L n v}{2L} \approx 3 \times 10^{-11}$ smaller. Since $v$ is extremely small, the propagated error in the time difference and thus the zero point of the HOM dip would be negligible. 
\subsection{Coherence length of photons}
\label{coherence}
We note that we can extend the time of rotation by increasing windings and therefore $L'=(2N+1) \pi R \sqrt{1-v^2}$ where $N$ is the number of windings. Thus the visibility for very slow rotation becomes $\mathcal{V}=1-P_{Gauss}/P_{Max}=e^{-\frac{\sigma^2 \Delta t^2}{2}}=e^{-8 \frac{v^2}{1-v^2}(2N+1)^2 \pi^2 R^2 \sigma^2}$. Thus far, we have been assuming timed pulses of light. Alternatively, the source of down-converted photons could be continuous characterized by a coherence length.
Note that the coherence length is defined as $\Delta x=\frac{2 \pi}{\sigma}$ where $\sigma$ is the width in frequency space. 
We consider parameters of $v=\frac{\Omega R}{c}$ where $R=20$ $cm$ and $\Omega=2 \pi$ $Hz$ with optical fibre of length $L'=10$ $km$. Therefore, the coherence length needed for significant visibility loss is $\Delta x= 4 \pi L' \frac{\Omega R}{c} \approx 500$ $\mu m$ which is easily coherence length of down-converted photons. In the next section, we will consider a quasi-continuous source and the effect of dispersion on the phase.

 

\subsection{Dispersion cancellation}
We've seen that the phase velocity isn't affected by the relativity of the co- and counter- propagating light for fibre of equal length. However, we now consider the full treatment of the effects of dispersive broadening and the group velocity in Appendix \ref{appendixa}. 

We find the usual phase but with a correction 
\begin{equation}
\Delta \phi_{v_g}=2\omega' L(\frac{1}{v_g^-}-\frac{1}{v_g^+})\approx \frac{4 \omega_0 v L}{1-v^2} (1+\frac{n'(k_0) v}{1-v^2}).
\end{equation}
Thus, as expected when the index of refraction is a constant $v_g=v_p$ and $n'(k_0)=0$ we obtain the usual phase shift with the cancellation of the refractive index. The correction is negligible for an $L=10$ $km$ long fibre at extremely slow rotation of $\Omega=2 \pi$ $Hz$ and radius $R=20$ $cm$. For optical fibre $n'(k_0)=-10^{-9}$ $m$, which is much less than unity and the correction term is negligibly small $5 \times 10^{-18}$.

We consider the frequency distribution $|f(\omega')|^2=(\frac{1}{\pi \sigma^2})^{1/2} \exp(-\frac{\omega'^2}{\sigma^2})$. The probability is therefore
\begin{equation}
\begin{split}
P_c&=\int{d\omega' |f(\omega')|^2 (1-\cos{(2\omega' \Delta \alpha L)}}) \\
&=\frac{1}{2}(1-\exp{(-4 \sigma^2 (\frac{v L}{1-v^2} (1+\frac{n'(k_0) v}{1-v^2}))^2 )}).
\end{split}
\end{equation}

Thus, we have the visibility with the corrected time delay from the group velocity. Ultimately, the effect of dispersion due to the material is cancelled out, and the group velocity effect is far too small ($n'(k)\approx-10^{-9}$ in silicon fibre) for the parameters suggested. For unequal lengths of the optical fibre, the phase term due to the index of refraction can be cancelled out using a time delay.

\section{Conclusion}
We have shown how to measure the visibility loss of interfering paths of a photon travelling in the Kerr metric. The effect of the non-trivial Kerr space-time manifests as a classical phase but the effect of time dilation in this metric can be measured by the visibility loss of single-photon interference. We can directly measure the time delay and loss in the visibility of quantum interference using a single-photon Sagnac interferometer in optical fibre. We also find that dispersion in the optical fibre cancels out and the difference in group velocity is negligible for the parameters considered. Based on our in-depth analysis, we draw parallels between single-photon interference in a rotating reference frame and the Kerr metric. This provides a recipe for near term experiments which could reproduce non-trivial results from the Kerr metric in a rotating system. Ultimately, one would like to observe such effects at the interface between quantum mechanics and non-trivial space-times.

{\it Acknowledgements}. We thank Matthias Fink and Rupert Ursin from the Institute for Quantum Optics and Quantum Information of the Austrian Academy of Sciences for discussions of potential experimental set ups. This work was supported in part by the Australian Research Council Centre of Excellence for Quantum Computation and Communication Technology (Project No. CE110001027) and financial support by an Australian Government Research Training Program Scholarship. We acknowledge that similar work in Ref. \cite{rest} was published at the same time as thesis submission containing some of the research presented here \cite{kishthesis}.

\section*{Data Availability}
The data that supports the findings of this study are available within the article.

\paragraph*{}
\section*{Conflict of Interest}
The authors have no conflicts to disclose.

\appendix
\section{Dispersion cancellation}
\label{appendixa}
In a moving medium, the phase velocity of light for co- and counter- propagating light is given by
\begin{equation}
v_{p\pm}=\frac{\omega}{k}=\frac{1-v^2}{n(k)\mp v},
\end{equation}
and the group velocity is given by
\begin{equation}
v_{g \pm}=\frac{1}{\alpha_{\pm}}=v_{p \pm}-\frac{d n(k)}{dk} \frac{1-v^2}{(n(k) \mp v)^2}=v_{p \pm} (1-\frac{n'(k)}{n(k)\mp v}),
\end{equation}
where $\alpha_{\pm}=\frac{d k_\pm}{d\omega_\pm}$ is the inverse group velocity.
The second order effect responsible for broadening is
\begin{equation}
\begin{split}
\frac{d^2\omega_\pm}{dk_{\pm}^2}&=\frac{1}{\beta_{\pm}}=v_{g \pm}-v_{p\pm}-\frac{d^2 n(k)}{dk^2} \frac{1-v^2}{(n(k)\mp v)^2}\\
&+2(\frac{d n(k)}{dk})^2 \frac{1-v^2}{(n(k)\mp v)^3},
\end{split}
\end{equation}
where the group velocity dispersion (GVD) is defined as $\beta=\frac{d^2 k}{d \omega^2}$. 
For example, for fused silica, the index of refraction is approximately linear with the wavelength. Thus as a function of $k$, $n(k)=\frac{100000}{k}+1.44$ around the wavenumber $k_0=8\times 10^6$ $m^{-1}$. Thus $n(k_0)=1.453$ and the derivative is $\frac{dn(k_0)}{dk}=-\frac{10^5}{k_0^2}=-1.6 \times 10^{-9}$ $m$. The second derivative is $\frac{d^2n(k_0)}{dk^2}=\frac{2 \times 10^5}{k_0^3}=4 \times 10^{-16}$ $m^2$. Thus $\frac{d^2\omega}{d k^2}\approx 1 \times 10^{-9}$ $m^2/s^2$.

We consider the following quasi continuous wave input state of down-converted light
\begin{equation}
\ket{\psi}=\int{ d\omega' f(\omega') \ket{\omega_0+\omega'}_A \ket{\omega_0-\omega'}_B}.
\end{equation}

After passing through the beamsplitter the modes in the two arms $1$ and $2$ are
\begin{eqnarray}
a_1(\omega_1)&=&\frac{i}{\sqrt{2}} a_A(\omega_1) e^{i k_A(\omega_1) L}+\frac{1}{\sqrt{2}} a_B (\omega_1) e^{i k_B(\omega_1) L}, \\
a_2(\omega_2)&=&\frac{i}{\sqrt{2}} a_B(\omega_2)e^{i k_B(\omega_2) L}+\frac{1}{\sqrt{2}} a_A (\omega_2) e^{i k_A(\omega_2) L},
\end{eqnarray}
where $k_A=k_0+\alpha_+ (\omega-\omega_0)+\beta_+ (\omega-\omega_0)^2$ is the wavenumber of the co-propagating light in the laboratory reference frame.
Similarly, $k_B=k_0+\alpha_- (\omega-\omega_0)+\beta_- (\omega-\omega_0)^2$ is the counter- propagating light. 

As in Ref. \cite{steinberg}, we assume a gate window time that is much larger than the dispersive broadening. This implies that cross terms of annihilation operators at different frequencies disappear for sufficiently long detector time scales. Thus the probability $P_c$ is

\begin{equation}
P_c \propto \int{d\omega_1} \int{d\omega_2} \braket{\psi|a^\dagger_1(\omega_1) a_2^\dagger (\omega_2) a_1 (\omega_1) a_2 (\omega_2) |\psi}.
\end{equation}

As it turns out, the phase term $\beta(\omega-\omega_0)^2$ acquired is the same in both interferometer arms due to the condition that $\omega_p=\omega_A+\omega_B$. The kernel is
\begin{equation}
\begin{split}
&\braket{\psi|a^\dagger_1(\omega_1) a_2^\dagger (\omega_2) a_1 (\omega_1) a_2 (\omega_2) |\psi} \\
=&|\frac{1}{2} \delta(\omega_p-\omega_1-\omega_2) f(\omega') [e^{i k_B (\omega_1) L+i k_A(\omega_2) L}\\
&-e^{-i k_B(\omega_2) L-i k_A(\omega_1) L}]|^2 \\
&=|\frac{1}{2} \delta(\omega_p-\omega_1-\omega_2) f(\omega') [e^{i(\Delta \alpha \omega'+\beta' \omega'^2)L}\\
& -e^{i (-\Delta \alpha \omega'+\beta' \omega'^2)L}]|^2,
\end{split}
\end{equation} 

where $\omega'=\omega_1-\omega_0=\omega_0-\omega_2$, $\Delta \alpha=\alpha_+ -\alpha_-$ and $\beta'=\beta_++\beta_-$. Evaluating the absolute squares
\begin{equation}
P_c=\int{d\omega' |f(\omega')|^2 (1-\cos{(2\omega' \Delta \alpha L))}}.
\end{equation}
\end{document}